\documentclass{elsart}
\usepackage[T1]{fontenc}
\usepackage[latin1]{inputenc}
\usepackage{graphicx}

\providecommand{\tabularnewline}{\\}

\usepackage{amssymb}

\begin{document}

\begin{frontmatter}
	\title{Universal Scaling in Saddle-Node Bifurcation Cascades (I) }

	\author[addr1,addr2]{Jesús San-Martín}
	\address[addr1]{Departamento de Matemática Aplicada, E.U.I.T.I, Universidad Politécnica de Madrid. Ronda de Valencia 3, 28012 Madrid Spain}
	\address[addr2]{Departamento de Física Matemática y Fluidos, U.N.E.D. Senda
del Rey 9, 28040 Madrid Spain}
	\ead{jsm@dfmf.uned.es}

	\begin{abstract}
A saddle-node bifurcation cascade is studied in the logistic equation,
whose bifurcation points follow an expression formally identical to
the one given by Feigenbaum for period doubling cascade. The Feigenbaum
equation is generalized because it rules several objects, which do
not have to be orbits. The outcome is that an attractor of attractors
appears, and information about the birth, death and scaling of windows
is obtained.
	\end{abstract}

	\begin{keyword}
Saddle-Node bifurcation cascade. Attractor of attractors. Generalized
Feigenbaum equation. Scaling windows. Scaling Myrberg-Feigenbaum points.
Bifurcation rigidity.
	\end{keyword}

\end{frontmatter}

\section{Introduction}

The study of the logistic equation

\begin{equation}
x_{n+1}=f(x_{n})=rx_{n}(1-x_{n})\,\,\,\,\,\,0\leq r\leq4\label{eq:recurrencia}\end{equation}
bears relevance because it is a simple expression with a numerical
and theoretical easy use that at the same time shows universal and
varied behaviors. It is enough to mention the patterns of periodic
orbit, found by Metropolis, Stein and Stein \cite{Metropolis73},
or the Feigenbaum cascade \cite{Feigenbaum78,Feigenbaum79}, the accumulating
point of which was discovered by Myrberg \cite{Myrberg63}.

As all quadratic maps are topologically conjugate \cite{Milnor88}
we can just focus the study in one of them. On the other hand, the
iterated one-dimensional maps, under rather general conditions, are
nearly quadratic if they are renormalized \cite{Gukenheimer87}. As
a consequence, the work done with the logistic equation can be generalized
to other maps. Therefore, any progress that makes clear the hierarchies
of bifurcations and the structure of this equation helps to the extension
of our knowledge of dynamical systems and the understanding of other
associated phenomena.

Within the study of the logistic equation the identification and organization
of orbits is outstanding, that is, its organization and hierarchies
in more detailed structures which are self-similar. It is also relevant
to establish in which order they are created, and from which kind
of bifurcation, as well as, for which parameter values the orbits
prevail \cite{Sharkovsky64,Sharkovsky93,Li75,Zehg84}. In short, it
is a question of establishing the structure of periodic windows, a
question to which we want to contribute. 

The study of the logistic equation is also important in Physics, both
as a model \cite{Beck99} and because it helps to calculate magnitudes
which describe certain physics processes: Lyapunov exponents and topological
entropy. From an experimental point of view the spotting stable orbit
would allow the experimenter to locate himself in a stable region,
which is close to phenomena that he wants to observe: bifurcations
and several kinds of chaos. On the other hand, the periodic orbits
allow us a semiclassical approach of quantum mechanics \cite{Gutzwiller71}.
Therefore, the progress in the understanding, in its mathematical
aspect, of periodic orbits helps to the theoretical and experimental
development of Physics.

Lately, an area of the study of application of logistic equation has
focused on the distribution of periodic windows, their widths, their
locations and its bifurcation diagram \cite{Hunt97,Hunt99}, clearly
linked to central question, above mentioned, of establishing which
is the complex structure of those windows. Similarly, we want to contribute
with our work, in which we will show a saddle-node bifurcation cascade.
This is a sequence of saddle-node bifurcations (tangent bifurcation),
with a recurrence law identical to the one Feigenbaum found for period
doubling cascade, which we will allow us to a) calculate the parameter
values where the windows are born and die, and therefore their length,
b) the identification and organization of orbits in windows, c) spot
the accumulating point of a saddle-node bifurcation cascade, in which
there will be chaos.

The works by Feigenbaum \cite{Feigenbaum78,Feigenbaum79,Feigenbaum80}
and Hunt et al. \cite{Hunt99} are the cornerstones to get the above
mentioned points. The first one will be fundamental in the numerical,
geometrical and theoretical interpretation of the results, meanwhile
the second one will be so for the theoretical reconstruction of the
behavior of the logistic equation as shown in its numerical analysis. 

This paper is organized as follows:

In section II numerical results are shown. These results suggest that
in an arbitrary period-$j$ window, the birth of successive period-$q*2^{n}$
saddle-node orbits scale in the same way as in a Feigenbaum cascade.
In general, inside the same period-$j$ window, there will be several
saddle-node bifurcation cascades with the same basic period $q$.

In section III we get the theoretical explanation of these properties.
In the process we will find that not only the period doubling cascade
and saddle-node bifurcation cascade scale in the same way but also
that it is typical of other sequences which appear in the logistic
equation, such as accumulating points of Myrberg-Feigenbaum points.

In section IV, and following the work by Feigenbaum for the period
doubling cascade, we develop a similar work for saddle-node cascade
which ends up in Feigenbaum-Cvitanoviç equation. It allows to explain
the self-similarity of iterated functions and to generalize the results
to polynomials with a maximum other than quadratic.

The results of section IV allow to generalize the bifurcation rigidity
Principle \cite{Hunt99} which is done in section V.

Section VI shows a rule to get the symbolic sequences of the saddle-node
bifurcation cascade orbits, as well as a rule to order some of them.

In the conclusion section, some connections with a possible physical
model are shown, which will most likely make the phenomenon fit in
the framework described in this work.

\section{Numerical results}

In the logistic map the period-$2^{n}$ orbits stem from pitchfork
bifurcation cascade: Feigenbaum cascade. However, the period-$q=3,5,7,...,2n+1,..$
orbits stem from saddle-node bifurcations (it is also possible to
find saddle-orbits in which $q$ is even, but not power of two).

Let $f^{q}$ be the $q$-th iterated of $f$. When $f^{q}$, $f$
obtained from Eq. (1), has a saddle-node bifurcation $q$ saddle-node
fixed points are generated at the same time. The node and saddle points
pull away from each other when the control parameter is varied. Every
$q$ node point stands for an orbit, and each orbit will be, in turn,
subjected to a Feigenbaum cascade. As result of that period-$q*2^{n}$
orbits are born. However, there are period-$q*2^{n}$ orbits which
do not stem from a period-$q$ orbit displaying pitchfork bifurcation,
but from $f^{q*2^{n}}$ displaying saddle-node bifurcations.

\begin{figure}[h]

\caption{\label{cap:fig1}Period-$3$ Saddle-Node orbit in the canonical window,
for $r\simeq3.82842712$.}

\includegraphics[%
  width=0.70\textwidth,
  keepaspectratio,
  angle=-90]{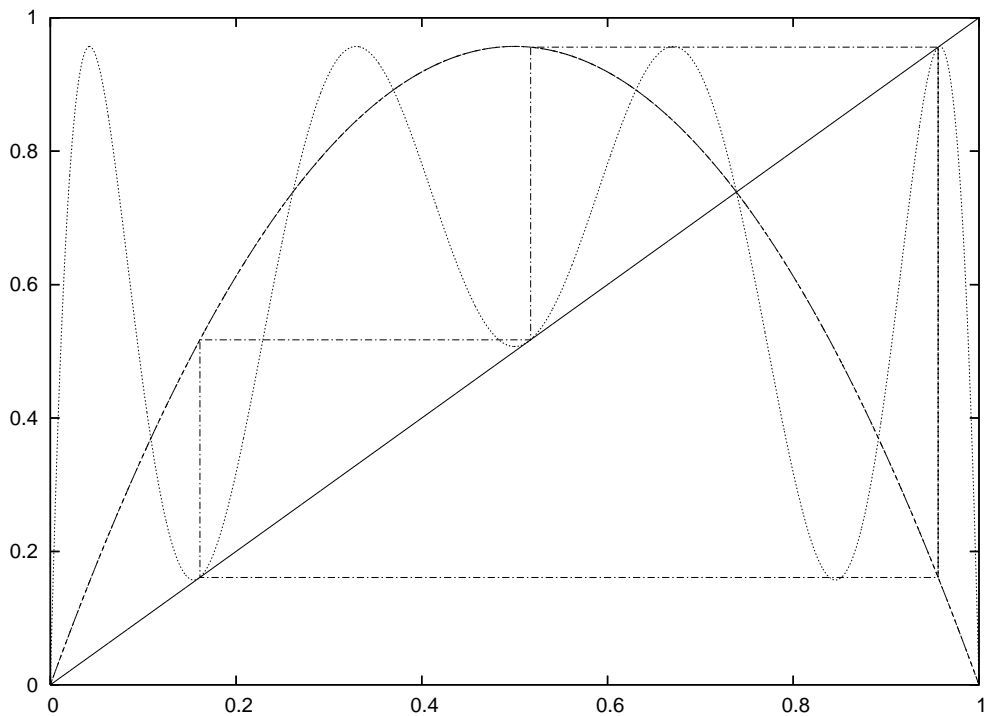}
\end{figure}

Whereas period-$q*2^{n}$ orbits duplicate their period because of
Feigenbaum cascade as the parameter $r$ is increased in Eq. (1),
contrary to what happens with saddle-node orbits. So, there is a period-$3$
saddle-node bifurcation at $r_{3}=1+\sqrt{8}$ \cite{Myrberg58} (see
Fig. \ref{cap:fig1}). If $r$ is increased, to get the point of pitchfork
bifurcation, a period-$3*2$ orbit is generated. A further increase
will generate a new pitchfork bifurcation and a period-$3*2^{2}$
orbit. A Feigenbaum cascade is taking place inside a period-$3$ window 

\begin{figure}[h]

\caption{\label{cap:fig2}Period-$3\cdot2$ Saddle-Node orbit in the canonical
window, for $r\simeq3.62655316$.}

\includegraphics[%
  width=0.70\textwidth,
  keepaspectratio,
  angle=-90]{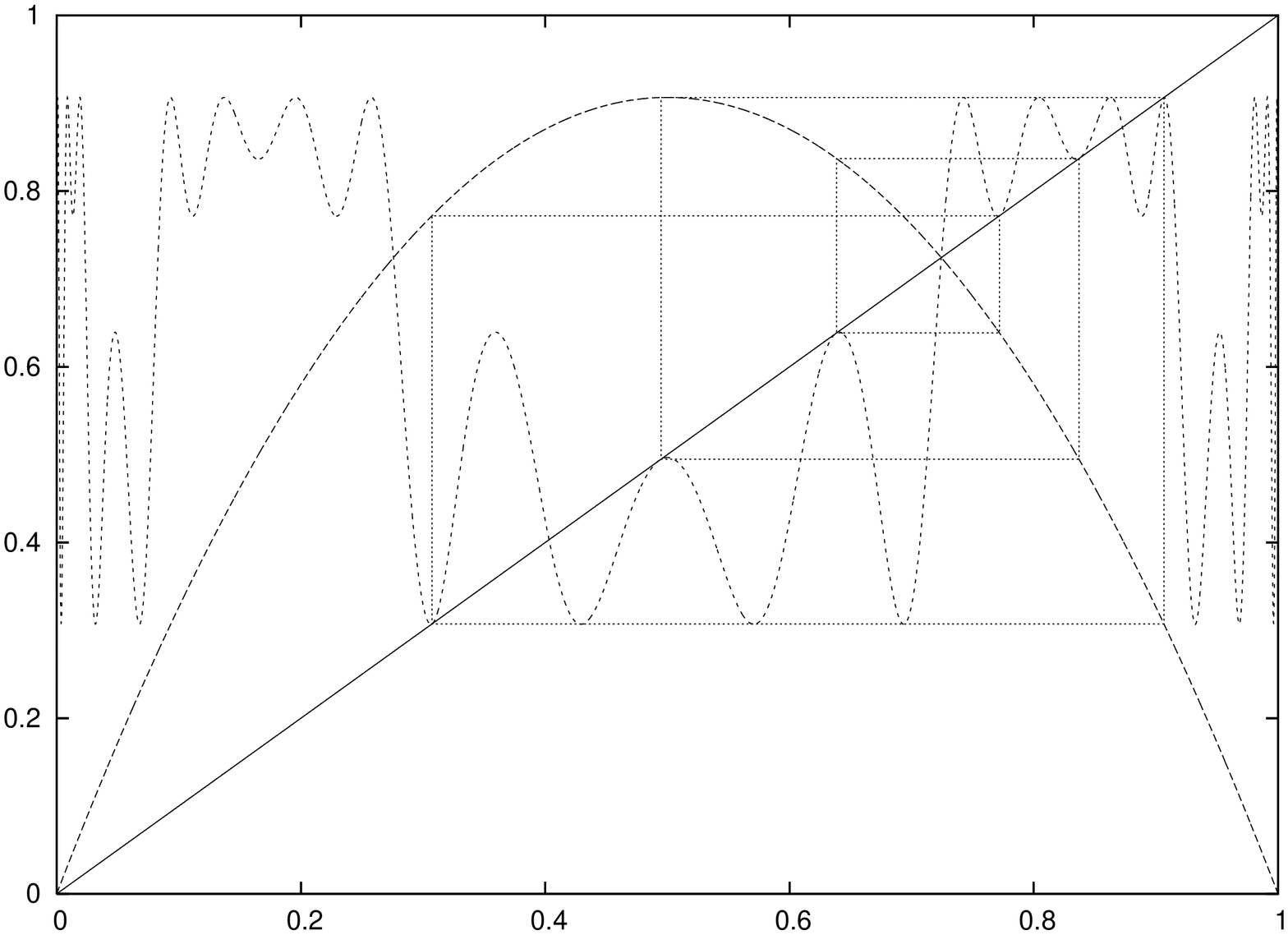}
\end{figure}

\begin{figure}[h]

\caption{\label{cap:fig3}Period-$3\cdot2^{2}$ Saddle-Node orbit in the canonical
window, for $r\simeq3.58202300$.}

\includegraphics[%
  width=0.70\textwidth,
  keepaspectratio,
  angle=-90]{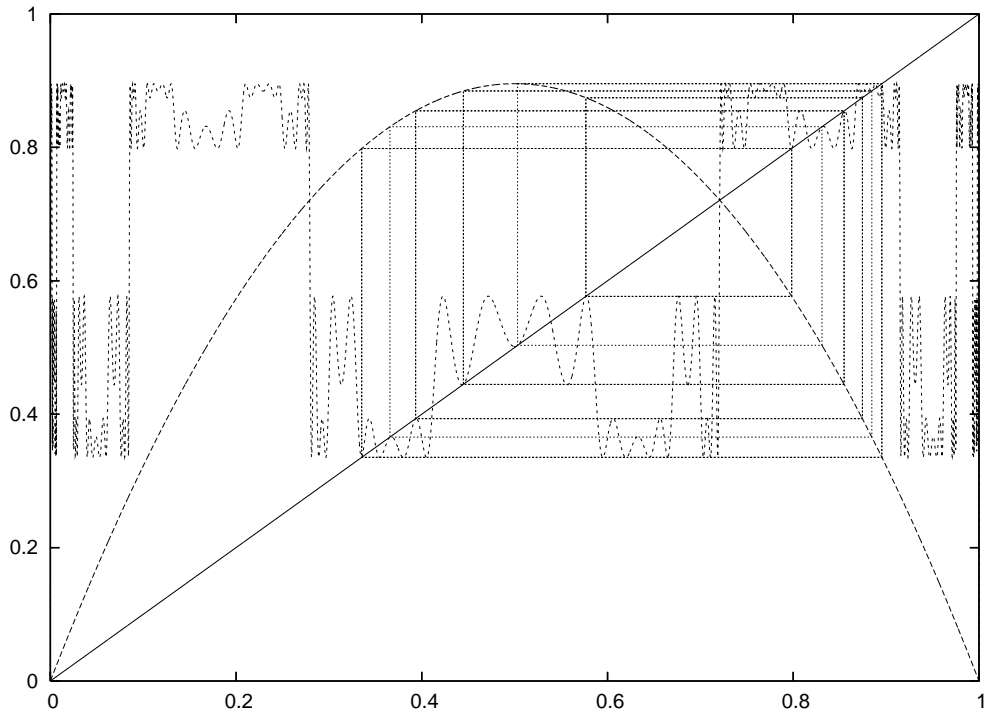}
\end{figure}

On the contrary if we decrease $r$ from $r_{3}=1+\sqrt{8}$ the period-$3$
orbit vanishes thus generating a saddle-node-$3*2$ period at $r_{3*2}\cong3.6265$
(see Fig. \ref{cap:fig2}). If $r$ is decreased further the saddle-node
period-$3*2$ orbit vanishes, and so a saddle-node period-$3*2^{2}$orbit
is generated at $r_{3*2^{2}}\cong3.5820$ (see Fig. \ref{cap:fig3}).
A period-$3,3*2,3*2^{2},..,3*2^{n},..$. saddle-node bifurcation cascade
takes place. Refer to Table \ref{cap:table1} for several birth saddle-node
orbit values. We can see that this saddle-node bifurcation cascade
is taking place at a canonical window, whereas pitchfork bifurcations
are taking place at a period-$3$window.

Following Feigenbaum \cite{Feigenbaum78} let's define 

\[
\delta_{3*2^{n}}=\frac{r_{3*2^{n+1}}-r_{3*2^{n}}}{r_{3*2^{n+2}}-r_{3*2^{n+1}}}\]
and

\[
\alpha_{3*2^{n}}=-\frac{d_{3*2^{n}}}{d_{3*2^{n+1}}}\]
where $r_{3*2^{n}}$ is the value of $r$ at which a period-$3*2^{n}$
orbit is created. And for this same saddle-node orbit we define $d_{3*2^{n}}$
as the distance from $x=\frac{1}{2}$ to the nearest saddle-node point.

In Table \ref{cap:table1} estimates of $\alpha_{n}$ and $\delta_{n}$
for several values of $n$ are shown, which apparently approach the
$\delta=4.66920160..$ and $\alpha=2.50290787..$. Feigenbaum constants.
Similar results are reached if we repeat the process for the saddle-node
period-$5,5*2,5*2^{2},..,5*2^{n},..$ orbit, as shown in Table \ref{cap:table2}.

The former two cases relate to saddle-node bifurcation cascade at
a canonical window, but the same kind of results are obtained working
with other windows. For instance, inside a period-$3$ window a saddle-node
bifurcation cascade can be generated, that is, period-$3,3*2,3*2^{2},..,3*2^{n},..$
orbits are created at this window. Nevertheless, it relates to period
$3*3,3*3*2,3*3*2^{2},..,3*3*2^{n},..$ orbits for the original map
Eq. (1). If this same cascade was generated at period-$5$window period-$5*3,5*3*2,5*3*2^{2},..,5*3*2^{n},..$
orbits would be created at a canonical window. In both cases, as shown
in Tables \ref{cap:table3} and \ref{cap:table4}, again we have the
same convergence to the $\delta$ Feigenbaum constant.

Let's emphasize that the saddle-node period-$5*3$ orbit is a saddle-node
period-$3$ orbit which displays at a period-$5$window; different
from a saddle-node period-$15$ displayed at a canonical window, and
different from a saddle-node period-$3*5$ orbit, as well, the latter
being born as a saddle-node period-$5$ orbit at a period-$3$ window
.

As well as numerical similarity with Feigenbaum's work, there is another
geometrical one. As shown in Figs. \ref{cap:fig1}, \ref{cap:fig2}
and \ref{cap:fig3} the graph of $f^{3}$(Fig. \ref{cap:fig1}) is
replicated in the neighborhood of $(\frac{1}{2},\frac{1}{2})$, for
any orbit of the period-$3,3*2,3*2^{2},..,3*2^{n},..$. saddle-node
bifurcation cascade. The same phenomenon can be noticed in other critical
points of $f^{2^{n}}$close to the line $y=x$. These geometrical
results apply not only to a $3,3*2,3*2^{2},..,3*2^{n},..$ cascade,
but also to $q,q*2,q*2^{2},..,q*2^{n},..$ $q\neq2^{m}$ cascades.

The geometrical, as well as the numerical, phenomenon apply not only
to a canonical window, as shown in Fig. \ref{cap:fig4}, where the
$3,3*2,3*2^{2},..,3*2^{n},..$ cascade is at the period-$5$window.

Summarizing, what the numerical results suggest is that for a given
saddle-node bifurcation, where a period-$q\neq2^{m}$ orbit is created,
there is a saddle-node bifurcation cascade where period-$q*2^{n}$
orbits are created. Let $r_{q*2^{n}}$ $n=0,1,2,3,...$ be the value
of $r$ at which a period-$q*2^{n}$ saddle-node bifurcation is created.
As $r$ is decreased from $r_{q}$ the different bifurcation values
appear as 

\[
...<r_{q*2^{n}}<...<r_{q*2}<r_{q}\]
 resulting in 

\begin{equation}
\delta=\lim_{n\rightarrow\infty}\delta_{q*2^{n}}=\lim_{n\rightarrow\infty}\frac{r_{q*2^{n+1}}-r_{q*2^{n}}}{r_{q*2^{n+2}}-r_{q*2^{n+1}}}\label{nueva 2}\end{equation}
and

\begin{equation}
\alpha=\lim_{n\rightarrow\infty}\alpha_{q*2^{n}}=\lim_{n\rightarrow\infty}\left(-\frac{d_{q*2^{n}}}{d_{q*2^{n+1}}}\right)\label{nueva 3}\end{equation}
where $\alpha,\delta$ are Feigenbaum's constants, and $d_{q*2^{n}}$
the distance from $x=\frac{1}{2}$ to the nearest saddle-point at
$r=r_{q*2^{n}}$.

As well as in the Feigenbaum cascade, when three consecutive values
$r_{q*2^{bin-3}},r_{q*2^{n-2}},r_{q*2^{n-1}}$are given the following
bifurcation value $r_{q*2^{n}}$ can be approximately predicted. This
result is important because the values where saddle-node bifurcation
appear are the values where period-$q*2^{n}$ windows are born. Therefore,
we can spot the birth of period-$q*2^{n}$ windows, once we have found
three consecutive bifurcation values. In addition, it is possible
to know how the lower endpoints of such periodic windows scale.

The Eq. (\ref{nueva 2}) and (\ref{nueva 3}) allow us to calculate
what it is happening in the non-primary windows, because each period-$q*2^{n}$
window, created in a saddle-node bifurcation, necessarily mimics the
canonical window.

\section{Generalization of Feigenbaum's formulas}

\subsection{Saddle-Node bifurcation cascade}

Following Feigenbaum work, let be the sequence

\begin{equation}
r_{1},r_{2},r_{3},.....,r_{n}.....\label{2 nueva 4}\end{equation}
made up by values of $r$, for which Eq. (\ref{eq:recurrencia}) shows
a pitchfork bifurcation. If the sequence is ordered in such a way
that a period-$j*2^{n}$ orbit is created at $r_{n}$, being $j$
the original orbit period, according to Feigenbaum \cite{Feigenbaum78}
there exist the limits

\begin{equation}
\delta=\lim_{n\rightarrow\infty}\frac{r_{n+1}-r_{n}}{r_{n+2}-r_{n+1}}=\lim_{n\rightarrow\infty}\delta_{n}\label{delta-F}\end{equation}

\[
r_{\infty}=\lim_{n\rightarrow\infty}r_{n}\]
From the former limits, the expression

\begin{equation}
r_{n+1}=\frac{1}{\delta}r_{n}+(1-\frac{1}{\delta})r_{\infty}\label{3 nueva 5}\end{equation}
follows, valid for $n>N$, being $N$ big enough.

For further development it is necessary to write Eq. (\ref{3 nueva 5})
in terms of the bounds of an interval and we can do so by using a
period-$j$ window originating from a saddle-node bifurcation. $r=r_{j,ini}$
is the lower bound of the window and $r=r_{j,end}$ the upper bound.
The window has a Myrberg-Feigenbaum point $r_{\infty,j}$, where a
period-$j*2^{n}$cascade finishes. Then, as Eq. (\ref{3 nueva 5})
there is an $r_{N,j}$ so that for every $r_{n,j}$ accomplishing
$r_{N,j}\leq r_{n,j}\leq r_{\infty,j}$ the equation

\begin{equation}
r_{n+1,j}=\frac{1}{\delta}r_{n,j}+(1-\frac{1}{\delta})r_{\infty,j}\label{4 nueva 6}\end{equation}
is accomplished.

With this posing, the lineal Eq. (\ref{3 nueva 5}) turns to lineal
Eq. (\ref{4 nueva 6}), which is accomplished in the subset $\left[r_{n,j};r_{\infty,j}\right]\subset\left[r_{j,ini};r_{j,end}\right]$.

We know that bifurcation points $r_{n,j}$ of doubling-period cascade
scale as Eq. (\ref{4 nueva 6}). We are going to prove that bifurcation
points of saddle-node bifurcation cascade, which have been encountered
numerically, escalate as Eq. (\ref{4 nueva 6}) too, that is

\[
r_{q*2^{n+1},SN,j}=\frac{1}{\delta}r_{q*2^{n},SN,j}+(1-\frac{1}{\delta})r_{\infty,j}\]
 at $r_{q*2^{n},SN,j}$ the function $f^{j}$ bifurcate to a saddle-node
period-$q*2^{n}$ orbit. 

To get this result the Eq. (\ref{4 nueva 6}) must extend to interval
$\left[r_{n,j};r_{q*2^{n},SN,j}\right]$. Accordingly let's take the
interval $\left[r_{n,j};r_{q*2^{n},SN,j}\right]\subset\left[r_{j,ini};r_{j,end}\right]$,
which holds the interval $\left[r_{n,j};r_{\infty,j}\right]$ where
Eq. (\ref{4 nueva 6}) is accomplished.

The key of proof is to chose two intervals $\left[r_{n,j};r_{q*2^{n},SN,j}\right]$
and$\left[r_{n+1,j};r_{q*2^{n+1},SN,j}\right]$, whose bounds are
bifurcation points of $f^{j}$. Both intervals will be related linearly.
The Eq. (\ref{4 nueva 6}) shows that the lower bound of an interval
is linearly transformed in the lower bound of the other interval.
By the uniqueness of linear transformation between both intervals,
the upper bounds are linearly transformed by Eq. (\ref{4 nueva 6}).

Starting from \cite{Yorke85}, Hunt \emph{et al.} \cite{Hunt99} demonstrated
the following principle:.

\emph{Principle 1: Let $X$ and $\Lambda$ be compact intervals. For
a typical $C^{3}$ family maps $f:X\times\Lambda\rightarrow X$ and
a typical superstable period $n$ orbit with critical point ($x_{0}$,$\lambda_{0}$),
there is a linear change of coordinates that conjugates $f^{n}$near
($x_{0}$,$\lambda_{0}$) to the quadratic family $y\longmapsto y^{2}-c$
in the square $[-2.5,\,2.5]\times[-2.5,\,2.5]$ to within an error
$\varepsilon$ in the $C^{2}$ norm, where $\varepsilon\rightarrow0$
as $n$ increases.}

Principle 1 shows that close to critical point the bifurcation diagram
for a window can be linearly transformed into a canonical bifurcation
diagram.

\begin{figure}[h]

\caption{\label{cap:fig4}Period-$3$ Saddle-Node orbit in the period-$5$
window, for $r\simeq3.74400300$. The shape of Fig. \ref{cap:fig1}
is reproduced five times; here is shown a magnification of the replica
near $(1/2,\,1/2)$.}

\includegraphics[%
  width=0.70\textwidth,
  keepaspectratio,
  angle=-90]{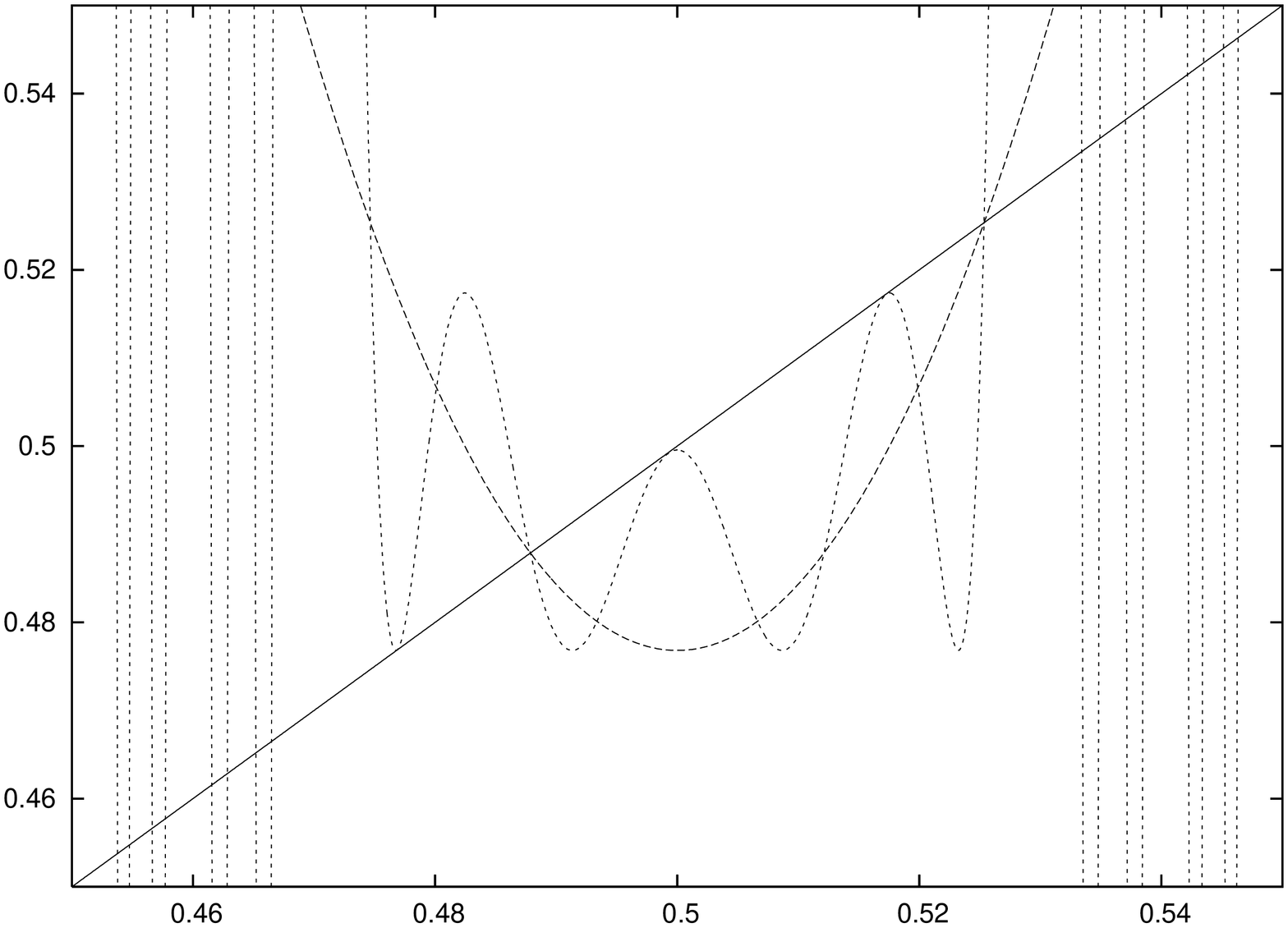}
\end{figure}

Let's see Fig. \ref{cap:fig4}. The first element of the $3*2^{m}$
saddle-node cascade is shown. In the neighborhood of $x=\frac{1}{2}$
of Fig. \ref{cap:fig4} the graph of $f^{3}$ is seen (Fig. \ref{cap:fig1}).
The graph of $f^{5}$ has been superimposed, which has a parabolic
shape in the neighborhood of $x=\frac{1}{2}$. If we expanded $f^{5}$
in the neighborhood of critical point $x=\frac{1}{2}$ it would turn
into a quadratic polynomial, which if iterated three times it would
look like a graph of $f^{3}$ as shown in Fig. \ref{cap:fig4}. The
same happens for all five critical points of $f^{5}$, each one of
which would have a graph of $f^{3}$ tangent to line $y=x,$and the
result would be a period-$3*5=15$ orbit in the canonical window.

Principle 1 suggests that the bifurcation diagram (of quadratic approximation
of $f^{5}$ close to critical point) is linearly transformed into
canonical bifurcation diagram. This linear transformation turns value
$r$ , which accounts for graph of $f^{3}$in Fig. \ref{cap:fig4},
into value $r$ which accounts for graph of $f^{3}$ in Fig. \ref{cap:fig1}.

As said before we formulate the following principle:

\textbf{Principle}.- In a period-$j$ window, saddle-node bifurcation
cascades scale by

\[
r_{q*2^{n+1},SN,j}=\frac{1}{\delta}r_{q*2^{n},SN,j}+(1-\frac{1}{\delta})r_{\infty,j}\]

Proof.

If the bifurcation diagram of $f^{j}$in $\left[r_{n,j};r_{q*2^{n},SN,j}\right]$
and the $f^{2*j}$in $\left[r_{n,2*j};r_{q*2^{n},SN,2*j}\right]$
are considered, according to Principle 1 there will be

i) A linear transformation that maps bifurcation diagram of $f^{j}$
in $\left[r_{n,j};r_{q*2^{n},SN,j}\right]$ onto canonical bifurcation
diagram of $f$ in $\left[r_{n,1};r_{q*2^{n},SN,1}\right]$, let this
be $h_{j}$. At $r_{n,1}$ the $n$th pitchfork bifurcation of $f$
occurs. At $r_{q*2^{n},SN,1}$a period-$q*2^{n}$ saddle-node bifurcation
of $f$ occurs.

ii) a linear transformation that maps bifurcation diagram of $f^{2*j}$in
$\left[r_{n,2*j};r_{q*2^{n},SN,2*j}\right]$ onto bifurcation diagram
of $f$ in $\left[r_{n,1};r_{q*2^{n},SN,1}\right]$, let this be $h_{2j}$.

Therefore, in the parameter space, there will be a linear transformation
$h_{2j}^{-1}\circ h_{j}$ that maps the bifurcation diagram of $f^{j}$
in $\left[r_{n,j};r_{q*2^{n},SN,j}\right]$ onto bifurcation diagram
of $f^{2*j}$ in $\left[r_{n,2*j};r_{q*2^{n},SN,2*j}\right]$. As
the bifurcation diagram of $f^{2*j}$in $\left[r_{n,2*j};r_{q*2^{n},SN,2*j}\right]$
coincides with bifurcation diagram of $f^{j}$ $\left[r_{n+1,j};r_{q*2^{n+1},SN,j}\right]$,
the result is that the linear transformation $h_{2j}^{-1}\circ h_{j}$
maps the bifurcation diagram of $f^{j}$ of $\left[r_{n,j};r_{q*2^{n},SN,j}\right]$
onto the bifurcation diagram of $f^{j}$in $\left[r_{n+1,j};r_{q*2^{n+1},SN,j}\right]$.

Let's notice how Eq. (\ref{4 nueva 6}) is a linear transformation
that maps the interval $\left[r_{n,j};r_{\infty,j}\right]$ onto $\left[r_{n+1,j};r_{\infty,j}\right]$.
As the linear transformation is unique, the linear transformation
$h_{2j}^{-1}\circ h_{j}$ within the limits of $\left[r_{n,j};r_{\infty,j}\right]\subset\left[r_{n,j};r_{q*2^{n},SN,j}\right]$
must coincide with Eq. (\ref{4 nueva 6}), i.e., $r_{n+1,j}=h_{2j}^{-1}\circ h_{j}(r_{n,j})=\frac{1}{\delta}r_{n,j}+(1-\frac{1}{\delta})r_{\infty,j}$.
However, the linear transformation is accomplished on the whole interval
$\left[r_{n,j};r_{q*2^{n},SN,j}\right]$and not only on the $\left[r_{n,j};r_{\infty,j}\right]\subset\left[r_{n,j};r_{q*2^{n},SN,j}\right]$
where a pitchfork bifurcation cascade occurs. By applying the same
lineal transformation to upper bound of interval the expression

\begin{equation}
r_{q*2^{n+1},SN,j}=\frac{1}{\delta}r_{q*2^{n},SN,j}+(1-\frac{1}{\delta})r_{\infty,j}\label{5 nueva 7}\end{equation}
results.

For that reason it is demonstrated that bifurcation parameter values
in saddle-node bifurcation cascade are governed by the same relation
that governs Feigenbaum cascade as shown in Eq. (\ref{3 nueva 5}).
Accordingly, the convergence on $\delta$of Feigenbaum

\[
\delta=\lim_{n\rightarrow\infty}\delta_{q*2^{n}}=\lim_{n\rightarrow\infty}\frac{r_{q*2^{n+1}}-r_{q*2^{n}}}{r_{q*2^{n+2}}-r_{q*2^{n+1}}}\]
suggested by numerical results is satisfied.

Obviously a saddle-node bifurcation cascade in a canonical window
escalates in the same way as in a period-$j$ window, as a result
of the linear relation between the bifurcation diagram of one window
and the other according to Principle 1.

\subsection{Non-uniqueness of Saddle-Node bifurcation cascade}

Most of the times, $f^{j}$ will have saddle-node bifurcations not
only for a single value of $r$ but for several (although $f^{3}$
has only one saddle-node bifurcation). Therefore, the interval $\left[r_{n,j};r_{q*2^{n},SN,j}\right]$
used to obtain the Eq. (\ref{5 nueva 7}) will not be unique, as there
will be different values $r_{q*2^{n},SN,j}$, with fixed $q,n,j$,
for which $f^{j}$ has saddle-node bifurcation. That means, there
will be different values $r_{q*2^{n},SN,j}$, with fixed $q,n,j$,
accomplishing the same Eq. (\ref{5 nueva 7}). Therefore, the saddle-node
bifurcation cascade that escalates as Eq. (\ref{5 nueva 7}) will
not be unique either. Although saddle-node orbits will have the same
period $q*2^{n}$.

For instance, $f^{6}$ shows three different $6*2^{n}$ saddle-node
bifurcation cascades, and none of them is related with the $3*2^{n+1}$
saddle-node cascade, although all of them have orbits with the same
periods. As shown in Table \ref{cap:table1}, it is obvious that $f^{6}$
has a saddle-node bifurcation at $r\cong3.6265$, but it must not
be understood that a $6*2^{n}$ saddle-node bifurcation cascade starts
at this point, but it must be considered as the second element of
the $3*2^{n}$ cascade, since the graph of $f^{3}$ appears in the
neighborhood of the critical points of $f^{2}$nearest to line $y=x$
(see Fig. \ref{cap:fig2}). The following element of the $3*2^{n}$cascade
occurs at $r\cong3.5820$ (see Fig. \ref{cap:fig3}), where the same
figure is repeated $2^{2}$ times. However, $f^{6}$ has saddle-node
bifurcation at $r_{1}\cong3.9375$, $r_{2}\cong3.9779$ and $r_{3}\cong3.9976$,
and it is impossible for $f^{3}$ to have saddle-node bifurcation
at these points. In each of all three former points a $6*2^{n}$ saddle-node
cascade starts. The following element of the respective cascades must
reproduce the graph of $f^{6}$ and duplicate its period, which occur
at $r_{a}\cong3.6552$, $r_{b}\cong3.6684$ and $r_{c}\cong3.6767$
respectively. Obviously, for these three new values $f^{12}$ has
a saddle-node bifurcation, but as said before, it must not be understood
as the beginning of a $12*2^{n}$ cascade, because the graph of $f^{6}$
is repeated twice. In Fig. \ref{cap:fig6} and \ref{cap:fig7} two
elements of $6*2^{n}$ cascade are shown, corresponding to $r_{2}$
and $r_{b}$.

\subsection{Other objects scaling with Feigenbaum's relation}

The reasoning followed to show the saddle-node bifurcation cascade
scaling can be used with other {}``objects'' . To do so, such {}``objects''
must belong to a sequence, and later build intervals with the elements
of the sequence in a similar way to the proof of a saddle-node bifurcation
cascade.

We can probe it with Myrberg-Feigenbaum points. To do so, let us consider
the period-$q*2^{n}$window $n=0,1,2,..$ , born of a primary saddle-node
orbit. Each window will have a Myrberg-Feigenbaum point $r_{\infty,q*2^{n}}$,
where the Feigenbaum cascade finishes. 
Let us build the sequence

\begin{equation}
r_{\infty,q,j},r_{\infty,q*2,j},...,r_{\infty,q*2^{n},j}\label{nueva 8}\end{equation}
that represents the parameter values of Eq. (\ref{eq:recurrencia}),
in which every one is the end of a Feigenbaum cascade in period-$q,q*2,q*2^{2},..,q*2^{n},..$
windows respectively, that is , Myrberg-Feigenbaum points. Period-$q,q*2,q*2^{2},..,q*2^{n},..$
windows are inside period-$j$ window, because of this the subindex
$j$ in Eq. (\ref{nueva 8}). This sequence is vital for the proof.

In proof of the saddle-node cascade convergence let $\left[r_{n,j};r_{q*2^{n},SN,j}\right]$
and $\left[r_{n,2*j};r_{q*2^{n},SN,2*j}\right]$ be substituted by$\left[r_{n,j};r_{\infty,q*2^{n},j}\right]$
and $\left[r_{n,2*j};r_{\infty,q*2^{n},2*j}\right]$. Initially $r_{q*2^{n},SN,j}$
used to indicate the value of parameter for which $f^{j}$ had a period-$q*2^{n}$
saddle-node orbit. Such orbit will have a Myrberg-Feigenbaum at some
value of $r$, which we represent by $r_{\infty,q*2^{n},j}$. This
is the value that appears in the new intervals, and also in the sequence
(\ref{nueva 8}). 

In order to demonstrate that the Myrberg-Feigenbaum points are governed
by 

\begin{equation}
r_{\infty,q*2^{n+1},j}=\frac{1}{\delta}r_{\infty,q*2^{n},j}+(1-\frac{1}{\delta})r_{\infty,j}\label{6 nueva 9}\end{equation}
$n>N$ , being $N$ large enough, we simply need to rebuild the saddle-node
cascade demonstration in the initial way, but with new intervals.

Let's expound what Eq. (\ref{6 nueva 9}) means. There will be saddle-node
period-$q*2^{n}$orbits inside a period-$j$ window born of a period-$j$
saddle-node orbit. After the birth of these orbits at $r_{q*2^{n},SN,j}$a
Feigenbaum cascade occurs, which will finish at Myrberg-Feigenbaum
point $r_{\infty,q*2^{n},j}$. Then, the Eq. (\ref{5 nueva 7}), which
interrelates birth-point of period-$q*2^{n}$ orbits, is identical
to the equation which interrelates the end-points where doubling-period
cascades of those same orbits finish.

The Eq. (\ref{6 nueva 9}) brings more information about Myrberg-Feigenbaum
points. It is well known that the Myrberg-Feigenbaum point $r_{\infty,j}$
is the accumulation point of values $r$ where pitchfork bifurcations
occur, because a doubling-period cascade finishes at $r_{\infty,j}$.
However, the Myrberg-Feigenbaum point $r_{\infty,j}$is not only the
accumulation point of Feigenbaum cascade because the Eq. (\ref{6 nueva 9})
shows that $r_{\infty,j}$ is the accumulation point of Myrberg-Feigenbaum
points $r_{\infty,q*2^{n},j}$,that is, Myrberg-Feigenbaum point $r_{\infty,j}$
is the accumulation point of Myrberg-Feigenbaum points of the period-$q*2^{n}$
$q\neq2^{m}$ primary windows, which have been born inside period-$j$
windows. As Myrberg-Feigenbaum points are attractors what we are in
front of is an attractor of attractors of attractors of ...

The problem of the convergence has not yet been completed. Let's notice
that at $r_{\infty,j}$ not only a single sequence of Myrberg-Feigenbaum
points convergences, but infinite ones, because for each $q\neq2^{m}$
there is an associated sequence $q*2^{n}$ $n=0,1,2,3,...$, and for
each sequence the same happens. The whole process is repeated for
each value of $q\neq2^{m}$.

The reasoning followed with Myrberg-Feigenbaum points can be applied
to other {}``objects''. The reader cannot miss that if instead of
choosing Myrberg-Feigenbaum points $r_{\infty,q*2^{n},j}$, where
a Feigenbaum cascade finishes, we choose the point $r_{q*2^{n},end,j}$
where the window started at $r_{q*2^{n},SN,j}$finishes then equation 

\begin{equation}
r_{q*2^{n+1}end,j}=\frac{1}{\delta}r_{q*2^{n},end,j}+(1-\frac{1}{\delta})r_{\infty,j}\label{7 nueva 10}\end{equation}
$n>N$ , being $N$ large enough, will be the same. 

Notice that three demonstration have been carried out for three different
points of a same window respectively, that is, $r_{q*2^{n},SN,j}$,
$r_{\infty,q*2^{n},j}$ and $r_{q*2^{n},end,j}$, initial point, intermediate
point and final point.

From the equations found we can answer part of the questions initially
posed: {}``where windows begin and end'', the answers to which are
given by Eq. (\ref{5 nueva 7}) and (\ref{7 nueva 10}). From these
results the relation between the length $L_{n}$of the period-$q*2^{n}$
windows is obtained, because by subtracting Eq. (\ref{5 nueva 7})
and (\ref{7 nueva 10})

\[
L_{n+1=}r_{q*2^{n+1},end,,j}-r_{q*2^{n+1},SN,j}=\frac{1}{\delta}r_{q*2^{n},end,,j}-\frac{1}{\delta}r_{q*2^{n},SN,j}=\frac{1}{\delta}L_{n}\]
is obtained, that is,

\begin{equation}
\delta=\frac{L_{n}}{L_{n+1}}\label{8 nueva 11}\end{equation}

This equation simply means that the quotient of the length of two
intervals which are linearly transformed one onto other must be constant.
The constant $\delta$ must be the proportional factor in the lineal
map, or else, how the lengths of the two consecutive windows born
of saddle-node bifurcation cascade are contracted. For the same reason,
the equation

\begin{equation}
\delta=\frac{r_{q*2^{n},end,j}-r_{q*2^{n},M,j}}{r_{q*2^{n+1},end,j}-r_{q*2^{n+1},M,j}}\label{9 nueva 12}\end{equation}
applies.

Let's bear in mind that we are talking about period-$q*2^{n}$ windows,
which scales in the same period-$j$ window, and not about how a period-$j$
window scales with respect to a canonical window. Therefore there
is no conflict with other theoretical results \cite{Hunt97,Yorke85}.

\subsection{Generalized Feigenbaum's relation}

Eqs. (\ref{5 nueva 7}), (\ref{6 nueva 9}) and (\ref{7 nueva 10})
are identical to the Eq. (\ref{3 nueva 5}), and they are valid for
$r>r_{\infty}$, meanwhile Eq. (\ref{3 nueva 5}) is valid for $r<r_{\infty}$.
We can express all these results in only one equation and generalize
the Feigenbaum equation (Eq. (\ref{3 nueva 5})) , by writing 

\begin{equation}
r_{n+1,j}=\frac{1}{\delta}r_{n,j}+(1-\frac{1}{\delta})r_{\infty,j}\label{10 nueva 13}\end{equation}
where points $r_{n,j}$ show the elements of a sequence (pitchfork
bifurcation, saddle bifurcation, Myrberg-Feigenbaum point, supercycle
or any other element to which the proof of saddle-node bifurcation
applies).

The Eq. (\ref{10 nueva 13}), apart from being a generalization, informs
us about the order in which the different elements of the sequence
appear. The order will depend on the parameter value $r_{n,j}$ being
smaller or bigger than $r_{\infty,j}$.

From the Eq. (\ref{10 nueva 13}) we can deduce that if $r_{n,j}\prec r_{\infty,j}$
then $r_{n,j}\prec r_{n+1,j}\prec r_{\infty,j}$, on the contrary
if $r_{n,j}\succ r_{\infty,j}$ then $r_{\infty,j}\prec r_{n+1,j}\prec r_{n,j}$.
From which we deduce that if a bifurcation occurs for a value such
that $r_{n,j}\prec r_{\infty,j}$ then the next value of bifurcation
increases, in the same way as for pitchfork bifurcation. On the contrary,
when the bifurcation occurs for $r_{n,j}\succ r_{\infty,j}$ then
the next values of bifurcation decreases, in the same way as for saddle-node
bifurcation cascades.

In this ordering, part of these results are new. If we consider first-occurrence
orbits the former results turn out to be a particular case of a theorem
(Theorem 2.10 in \cite{Sharkovsky93}). However, the ordering shown
in this section deals with period-$j$ window, which does not have
to originate from first-occurrence period-$j$ windows. For this reason,
the ordering is not restricted to first-occurrence  orbits and therefore
it is not regarded in the Sharkovsky theorem.

On the other hand, the ordering that we have just shown is not restricted
only to orbits, as the Sharkovsky theorem, but it also applies to
Myrberg-Feigenbaum points and to other objects that are not orbits.
An ordering which was hidden as far as we know.

Therefore the Eq. (\ref{10 nueva 13}) enlarges the ordering to both
new orbits and other objects which are not orbits.

Let's summarize the conclusions obtained in this section for a period-$j$window:.

i) It is possible to spot the birth of a saddle-node period-$q*2^{n}$
orbit and therefore to fix the birth of period-$q*2^{n}$ windows
from Eq. (\ref{5 nueva 7}). This equation justifies the numerical
results shown in Tables \ref{cap:table1}, \ref{cap:table2}, \ref{cap:table3}
and \ref{cap:table4}, where the convergence to Feigenbaum $\delta$constant
is shown.

ii) Equally it spots where a period-$q*2^{n}$ window finishes from
Eq. (\ref{7 nueva 10})

iii) According to the former two points the scaling of windows related
to saddle-node bifurcation cascades is proved Eq. (\ref{8 nueva 11}).

iv) The Feigenbaum equation (Eq. (\ref{3 nueva 5})) has been generalized,
both in the range of the parameter validity and in the objects to
which is applied

v) In particular, the generalization of Eq. (\ref{3 nueva 5}) is
applied to Myrberg-Feigenbaum points, showing that these points behave
as an attractor of attractors.

vi) The generalization, applied to Myrberg-Feigenbaum points, allows
us to spot where the Feigenbaum cascade of period-$q*2^{n}$ window
finishes by means of Eq. (\ref{6 nueva 9}), or else, it shows the
relative position of the attractors in which doubling-period cascades
finish. 

vii) The ordering of the objects governed by Eq. (\ref{10 nueva 13})
is shown, and if they happen to be orbits they do not have to be first-occurrence
orbits.

In this way, we account for the question initially posed, about the
birth, end, scaling and structure of the orbits.

\section{Self-similarity and convergence to Feigenbaum's $\alpha$ constant}

\subsection{Self-similarity}

In the numerical results we anticipated that the graph of $f^{3}$(Fig.
\ref{cap:fig1}), when it has a saddle-node bifurcation, is repeated
in the neighborhood of point $(\frac{1}{2},\frac{1}{2})$ both $f^{3\cdot2}$
and $f^{3\cdot2^{2}}$, when $f^{3\cdot2}$ and $f^{3\cdot2^{2}}$have
also saddle-node bifurcation. Furthermore, the graph of $f^{3}$ in
the neighborhood of point $(\frac{1}{2},\frac{1}{2})$ appears contracted
by a factor $\frac{1}{\alpha}$, each time $f^{3}$is iterated with
itself, where $\alpha$is a Feigenbaum constant.

Let's justify the numerical results, as well as why similar figures
are shown at the $2^{n}$ critical points of $f^{2^{n}}$near line
$y=x$ (see Fig. \ref{cap:fig3}) These similar figures are like saddle-node
orbits trapped in the critical points, and we will use them for the
mathematical analysis.

A saddle-node period-$q*2^{n}$ orbit is obtained when the derivative
of $f^{q*2^{n}}$ takes value 1. This orbit is forced to a period
doubling process, a process that will occur when the derivative takes
value -1. Accordingly, the derivative vanishes in some middle point,
which corresponds to a supercycle. Therefore, we can associate each
saddle-node orbit with a supercycle.

Feigenbaum \cite{Feigenbaum78} introduced the quotient

\begin{equation}
\delta_{n}=\frac{R_{n+1}-R_{n}}{R_{n+2}-R_{n+1}}\label{nueva 14}\end{equation}
where $R_{n}$is the value for which Eq. (\ref{eq:recurrencia}) has
a period-$2^{n+1}$ supercycle in a period doubling cascade. The quotient
(Eq. (\ref{nueva 14})) approaches a$\delta$ Feigenbaum constant.

Let's follow Feigenbaum and define

\begin{equation}
\delta_{n}=\frac{r_{n+1}-r_{n}}{r_{n+2}-r_{n+1}}\label{nueva 15}\end{equation}
being $r_{n}$the value where the period-$q*2^{n}$ supercycle occurs.
Such supercycle is associated to a period-$q*2^{n}$ saddle-node orbit.
Then we will get the same results as Feigenbaum: the convergence of
the quotient (Eq. (\ref{nueva 15})) on $\delta$ constant (see Table
\ref{cap:table1}). This result leads us to the definition of a family
of functions similar to those created by Feigenbaum in his work

As for Eqs. (\ref{nueva 14}) and (\ref{nueva 15}), we have to emphasize
that we will work with supercycles associated to orbits born of saddle-node
bifurcations instead of supercycles associated to orbits born of pitchfork
bifurcation as Feigenbaum does. There is an important difference with
regard to the behavior of both kinds of orbits in the parameter space.
The orbits of period doubling cascade are adjacent in a connected
set: an interval of the parameter space. However, the saddle-node
period-$q*2^{n}$ orbits only exist in discrete points of the parameter
space. Besides, saddle-node orbits that do not belong to the same
cascade can be found between saddle-node period-$q*2^{n}$and period-
$q*2^{n+1}$ orbits, which proves that the latter orbits are not adjacent.
Let us mention other essential difference between both bifurcations:
the pitchfork bifurcation has a local character, whereas the Saddle-Node
bifurcation leads to a global qualitative change in the behavior of
the logistic equation.

Bearing in mind the previous explanation and following Feigenbaum,
we define ( see Eq. (35) of \cite{Feigenbaum80})

\[
g_{1,q}=\lim_{n\rightarrow\infty}\left(-\alpha\right)^{n}f_{R_{n+1,q}}^{2^{n}}\left[\frac{x}{\left(-\alpha\right)^{n}}\right]\]
where the supercycle associated to saddle-node period-$q*2^{n}$ orbit
is obtained at $R_{n,q}$, and $q$ is fixed.

From previous definition we define a family of functions (see Eq.
(39) of \cite{Feigenbaum80}) 

\begin{equation}
g_{i,q}=\lim_{n\rightarrow\infty}\left(-\alpha\right)^{n}f_{R_{n+i,q}}^{2^{n}}\left[\frac{x}{\left(-\alpha\right)^{n}}\right]\label{11 nueva 16}\end{equation}
where $q$ is fixed, which fulfills the next equation (see Eq. (42)
of \cite{Feigenbaum80})

\begin{equation}
g_{i-1,q}=\left(-\alpha\right)g_{i,q}\left[g_{i,q}\left(-\frac{x}{\alpha}\right)\right]\equiv Tg_{i,q}(x)\label{12 nueva 17}\end{equation}
Taking the limit in Eq. (\ref{11 nueva 16}) (see Eq. (40) of \cite{Feigenbaum80})
the function converges to a limiting function

\begin{equation}
g(x)=\lim_{i\rightarrow\infty}g_{i,q}\label{13 nueva 18}\end{equation}

This limit exists at a fixed value of $r$, the Myrberg-Feigenbaum
$r_{\infty}$, and it means that there is an accumulation point at
$r_{\infty}$ for a saddle-node $q*2^{n}$bifurcation cascade.

It follows from Eqs. (\ref{12 nueva 17}) and (\ref{13 nueva 18})
that

\begin{equation}
g(x)=-\alpha g\left[g\left(-\frac{x}{\alpha}\right)\right]\label{14 nueva 19}\end{equation}
(see Eq. (43) of \cite{Feigenbaum80})

It follows from Eq. (\ref{14 nueva 19}) that if $g(x)$ is the solution
of this equation then $g^{q}(x)$ is the solution of

\begin{equation}
g^{q}(x)=-\alpha g^{q}\left[g^{q}(-\frac{x}{\alpha})\right]\label{15 nueva 20}\end{equation}

This equation means that the iterated of $g^{q}$ is self-similarly
scaled by $\alpha$. Precisely, this is shown in Fig. \ref{cap:fig1},
\ref{cap:fig2} and \ref{cap:fig3}, in the neighborhood of $\left(\frac{1}{2},\frac{1}{2}\right)$:
the graph of $f^{3}$, close to a saddle-node bifurcation, is repeated
scaled by a constant. This constant is the nearer to $\alpha$ the
bigger $n$ is in $f^{3*2^{n}}$, if $f^{3*2^{n}}$ has a saddle-node
bifurcation (see Table \ref{cap:table1}). With this last reasoning
all numerical results shown in section 2 have been verified.

The Eq. (\ref{14 nueva 19}), that we have just found, is the Feigenbaum-Cvitanoviç
equation, which Feigenbaum utilized to show why function $f$ is repeated
in the neighborhood of $(\frac{1}{2},\frac{1}{2})$ in a period doubling
cascade. If so, we wonder, why the original work of Feigenbaum is
not utilized to justify the self-similarity of $f^{3}$in a saddle-node
bifurcation cascade. The reason lies in the fact that Feigenbaum reached
Eq. (\ref{14 nueva 19}) from a limit of functions associated with
pitchfork bifurcations, and accordingly its validity is not justified
when we work with functions associated with other kinds of bifurcations,
as is the case with saddle-node bifurcations.

This is not the only difference. It has been said that Eq. (\ref{14 nueva 19})
is the limit of Eq. (\ref{12 nueva 17}) only at $r=r_{\infty}$,
and Feigenbaum got this value on the left, i.e., with $r<r_{\infty}$
which is how a period doubling cascade occurs. However we have got
$r_{\infty}$ on the right, as saddle-node bifurcations occur for
$r>r_{\infty}$. That is, the Eq. (\ref{14 nueva 19}) can be reached
by limits of functions defined for both $r<r_{\infty}$and $r>r_{\infty}$,
and that explains why Eq. (\ref{10 nueva 13}) generalizes Eq. (\ref{3 nueva 5})
of Feigenbaum, as we will see later.

\subsection{Derivation of the generalized Feigenbaum formula.}

Feigenbaum \cite{Feigenbaum79} got the Eq. (\ref{3 nueva 5}) from
the equation

\begin{equation}
g_{i+1}(x)-g_{i}(x)=\delta^{-i}h(x)\,\,\,\, i\gg1\label{rel-F}\end{equation}
where $g_{i}(x)$ is the family of functions that we have utilized
to define the family $g_{i,q}$ given by Eq. (\ref{11 nueva 16}).
$g_{i}(x)$ and $g_{i,q}$ are formally identical, but $g_{i}(x)$
refers to pitchfork bifurcations and $g_{i,q}$ refers to saddle-node
bifurcations. In order to define well the Eq. (\ref{rel-F}) it is
necessary to point out what $h(x)$ and $\delta$ are, which are calculated
from

\begin{equation}
\delta h(x)=L_{g}h(x)\label{F-C-lineal}\end{equation}
where\begin{equation}
L_{g}h(x)=-\alpha\left\{ g'\left[g(\frac{-x}{\alpha})\right]h(\frac{-x}{\alpha})+h\left[g(\frac{-x}{\alpha})\right]\right\} \label{oper-linea}\end{equation}
where $\alpha$ and $g(x)$ are given by the Eq. (\ref{14 nueva 19}).

If we follow Feigenbaum, working with $g_{i,q}$ instead of $g_{i}$,
we will calculate an equation identical to Eq. (\ref{rel-F}), i.e., 

\begin{equation}
g_{i+1,q}(x)-g_{i,q}(x)=\delta^{-i}h(x)\,\,\,\, i\gg1\label{rel-Jesus}\end{equation}
where $g_{i,q}$ is the family of functions given by Eq. (\ref{11 nueva 16}),
whereas both $h(x)$ and $\delta$ are given by Eqs. (\ref{F-C-lineal})
and (\ref{oper-linea}). As both $g_{i}(x)$ and $g_{i,q}$ are identical
and both Eqs. (\ref{rel-F}) and (\ref{rel-Jesus}), which govern
the former functions, are also identical then the result has to be
the same: the Eq. (\ref{3 nueva 5}). Although, this time the equation
refers to a saddle-node bifurcations for $r>r_{\infty}$instead of
a pitchfork bifurcation cascade for $r<r_{\infty}$, as Feigenbaum
reached, that is, the Eq. (\ref{3 nueva 5}) is valid for both $r>r_{\infty}$
and $r<r_{\infty}$. The thing is that depending on the fact that
wether $r$ is smaller or bigger than $r_{\infty}$, the equation
is assigned to one bifurcation or another.

The result of the former explanation is that Eq. (\ref{10 nueva 13}),
which generalizes Eq. (\ref{3 nueva 5}), is obtained without the
bifurcation rigidity Principle of Hunt et al. \cite{Hunt99}, as it
was done in section 3.1, and it will be the starting point to generalize
the bifurcation rigidity Principle in section 5. 

Now we have obtained the equation which governs bifurcation points
in saddle-node bifurcation cascades. We must distinguish two cases:
saddle-node bifurcation cascades with the same or different basic
periods. In a $q,q*2,q*2^{2},..,q*2^{n},..$ $q\neq2^{m}$ saddle-node
bifurcation cascade $q$ will be named a basic period.

\subsubsection{Saddle-Node bifurcation cascades with the same basic period}

As shown in section 3,  there are different saddle-node bifurcation
cascades with the same basic period in the canonical window, which
are generated because the same period basic $q$ is produced for different
values of parameter $r$. Accordingly, there are different values
$R_{n,q}$ , with fixed $q$ and $n$, and therefore the family of
functions given by Eq. (\ref{11 nueva 16}) are not the only ones,
but there are as many families as there are values of $R_{n,q}$ (see
sections 3.2 and 4.1). Given that there are different $g_{i,q}$families
then there are different equations like Eq. (\ref{rel-Jesus}). As
Eq. (\ref{3 nueva 5}) is obtained from Eq. (\ref{rel-Jesus}) it
turns out that there will be one Eq. (\ref{3 nueva 5}) for each saddle-node
bifurcation cascade. All equations will be identical and all of them
will approach to $r_{\infty}$, because the limit given by Eq. (\ref{14 nueva 19})
exists only for $r_{\infty}$ which does not depend on value of the
$R_{n,q}$. What we have just found is something already known: for
each fixed period $q$ there are different saddle-node bifurcation
cascades, each one of which is governed by the Eq. (\ref{3 nueva 5}),
and all of them approach to $r_{\infty}$(see section 3.2).

\subsubsection{Saddle-Node bifurcation cascades with different basic period}

For each basic period there will be one saddle-node bifurcation cascade,
and therefore a family of functions $g_{i,q}$. Following the explanation
of section 4.2.1 there will be an equation like Eq. (\ref{3 nueva 5})
for each bifurcation cascade. All equations will be identical and
all of them will approach to $r_{\infty}$.

We have already found this result in section 3 (see Eq. (\ref{5 nueva 7})).

\subsection{Extension to non-canonical windows}

Until now we have only been working in the canonical window, but it
is not difficult to extend the results to any window. For instance,
if we had worked in a period-$p$ window then a $q\cdot2^{n}$ saddle-node
bifurcation cascade would have occurred inside it. It would be enough
to take into account that the $R_{n,q}$ of Eq. (\ref{11 nueva 16})
is the value for which the supercycle associated to a saddle-node
period-$q*2^{n}$ orbit occurs, but this time inside a period-$p$
window, so that the whole process is identical and the same universal
results are obtained. The difference would lie in the fact that the
approaching to the Eq. (\ref{14 nueva 19}) would be at a Myrberg-Feigenbaum
point $r_{\infty,p}$ of a period-$p$ window instead of at the Myrberg-Feigenbaum
point $r_{\infty}$ of the canonical window. Therefore, if $r_{\infty}$
were substituted by $r_{\infty,p}$ all equations valid in the canonical
window would remain valid in a period-$p$ window.

\subsection{Extension to functions with a non-quadratic maximum}

As we have generalized the results from a canonical window to an arbitrary
periodic window, one more generalization remains to be done: the results
are generalized to maps other than logistic ones.

In this paper, the work has been developed with Eq. (\ref{eq:recurrencia}),
which is topologically conjugate to 

\[
x_{n+1}=x_{n}^{2}-c\]

The whole work could have been developed with the function 

\[
f(x,c)=x^{2}-c\]
and it could have been repeated in an identical way with the family
of functions

\begin{equation}
f_{n}(x,c)=x^{2n}-c\,\,\,\,\, n=1,2,3,...\label{f-conj}\end{equation}

The sections 4.1 and 4.2 can be rewritten, step by step, defining
a new family of functions $g_{i,q}$, which are identical to the ones
given by Eq. (\ref{11 nueva 16}), but this time $f$ is replaced
by Eq. (\ref{f-conj}). This new family of functions applies to an
equation identical to Eq. (\ref{12 nueva 17}), and it results in
Eq. (\ref{14 nueva 19}), which has a solution \cite{Epstein89},
once the maximum of $g(x)$has been fixed as $x^{2n}$ , $n=1,2,3,...$,
although $\alpha$ and $\delta$ will be different, depending on the
kind of maximum.

As the universal behavior comes from Eq. (\ref{14 nueva 19}), and
this remains identical, it turns out that $g^{q}$is self-similar,
although we will have a different $g(x)$ depending on the kind of
maximum we choose. We will see how the graph of functions $x^{2n}$
is repeated in the saddle-node bifurcation cascade.

The results relative to saddle-node bifurcation cascades remain the
same, because these results came from Eqs. (\ref{rel-Jesus}), (\ref{F-C-lineal})
and (\ref{oper-linea}), which did not change after $f$ was replaced
by Eq. (\ref{f-conj}) in $g_{i,q}$. As everything is identical it
turns out that equations identical to Eq. (\ref{3 nueva 5}) will
be obtained, but the value of $\delta$ will depend on the kind of
maximum with which we are dealing with.

In short, in this section we have expound the convergence to Feigenbaum
constant $\alpha$ and the self-similarity of the function $f^{q}$
in the neighborhood of $\left(\frac{1}{2},\frac{1}{2}\right)$ as
it was shown in section 2. We have also laid the foundations for a
generalization of the Feigenbaum Eq. (\ref{3 nueva 5}) and, at last,
we have generalized these results to functions with maximum like $x^{2n}$
, $n=1,2,3,...$.

\section{Generalization of Bifurcation Rigidity Principle}

We have reached Eq. (\ref{3 nueva 5}) (see section 3.1) coming from
the bifurcation rigidity Principle \cite{Hunt99} for functions with
a quadratic maximum. The proof can be followed the other way round,
starting from Eq. (\ref{3 nueva 5}) and reaching the bifurcation
rigidity Principle. If we start from Eq. (\ref{3 nueva 5}), valid
for functions like Eq. (\ref{f-conj}), we will arrive to the bifurcation
rigidity Principle. This time it will be valid for functions like
Eq. (\ref{f-conj}) and not only for functions with quadratic maximum
as it was initialed expounded.

Let a function be like Eq. (\ref{f-conj}). According to section 4.4,
we know that both successive pitchfork and successive saddle-node
bifurcations are governed by an equation like

\begin{equation}
r_{n+1}=\frac{1}{\delta}r_{n}+(1-\frac{1}{\delta})r_{\infty}\label{20}\end{equation}

In period-$j$ window successive saddle-node bifurcations will be
governed by 

\begin{equation}
r_{q*2^{n+1},SN,j}=\frac{1}{\delta}r_{q*2^{n},SN,j}+(1-\frac{1}{\delta})r_{\infty,j}\label{21}\end{equation}
meanwhile in a canonical window the successive saddle-node bifurcations
will be governed by 

\begin{equation}
r_{q*2^{n+1},SN}=\frac{1}{\delta}r_{q*2^{n},SN}+(1-\frac{1}{\delta})r_{\infty}\label{22}\end{equation}

Accordingly there is a linear transformation between both saddle-node
bifurcation cascades given by

\begin{equation}
\frac{r_{q*2^{n+1},SN,j}-\frac{1}{\delta}r_{q*2^{n},SN,j}}{r_{q*2^{n+1},SN}-\frac{1}{\delta}r_{q*2^{n},SN}}=\frac{r_{\infty,j}}{r_{\infty}}\label{23}\end{equation}

For pitchfork bifurcations there is an identical equation

\begin{equation}
\frac{r_{n+1,j}-\frac{1}{\delta}r_{n,j}}{r_{n+1}-\frac{1}{\delta}r_{n}}=\frac{r_{\infty,j}}{r_{\infty}}\label{24}\end{equation}
where $n$ is related to $n$-th pitchfork bifurcation.

Let's take the intervals 

$\left[r_{n,j};r_{q*2^{n},SN,j}\right]$ and $\left[r_{n+1,j};r_{q*2^{n+1},SN,j}\right]$\\
in the period-$j$ window and the intervals 

$\left[r_{n};r_{q*2^{n},SN}\right]$ and $\left[r_{n+1};r_{q*2^{n+1},SN}\right]$\\
in the canonical window. If a linear transformation maps the interval
$\left[r_{n,j};r_{q*2^{n},SN,j}\right]$ onto the interval $\left[r_{n};r_{q*2^{n},SN}\right]$then
Eqs. (\ref{23}) and (\ref{24}) will force the interval $\left[r_{n+1,j};r_{q*2^{n+1},SN,j}\right]$
to be linearly transformed into the $\left[r_{n+1};r_{q*2^{n+1},SN}\right]$.
Accordingly the bifurcation diagrams of both window are linearly changed
as set by the bifurcation rigidity Principle, but this time the Principle
works with $x^{2n}$-maps and not only with quadratic maps. As the
Eq. (\ref{20}) is applied to $n\gg1$, the smaller the windows are,
the better the linear approximation to the transformation will be.

\section{Symbolic sequences and orbit ordering}

\textbf{Sequence principle}.- Let be a period-$p*(q*2^{n})$ saddle-node
orbit. The points of this orbit lie inside the $p*2^{n}$ neighborhoods
of the $p*2^{n}$ supercycle points of $f^{p*2^{n}}$, each one of
these neighborhoods having $q$ saddle-node points. Then the sequence
of the saddle-node orbit is obtained with the following process

i) Write the sequence of the orbit of the supercycle of $f^{p*2^{n}}$,
which will be like $CI_{1}.....I_{p*2^{n}-1}$, where $C$ indicates
the center and represents the point $x=\frac{1}{2}$ and $I_{i}$
, $i=1,...,p*2^{n}-1$ can take the values L (left) or R (right) with
respect to $C$.

ii) Write consecutively $q$ times the sequence obtained in the former
point i), getting a sequence like

$C_{1}I_{1}.....I_{p*2^{n}-1}C_{2}I_{1}.....I_{p*2^{n}-1}.........C_{q}I_{1}.....I_{p*2^{n}-1}$.

iii) Write the sequence of period-$q$ saddle-node orbit, that is,
the sequence of the saddle-node orbit of $f^{q}$.

iv) Calculate $\left(-1\right)^{n}$. If the result is negative then
conjugate the letters obtained in point iii) by means of L$\leftrightarrow$R.
Bear in mind that $n\in Z^{+}$.

v) Replace consecutively each letter $C_{1}C_{2}.....C_{q}$ of the
sequence obtained in ii) by the complete sequence obtained in iv),
keeping the order.

Proof

i) Let be period-$p*2^{n}$ supercycle at $r=r_{0}$, in the period-$p$
window, whose points are $\left\{ x_{1},...x_{p*2^{n}}\right\} $.
The order in which these points are visited determines the sequence
of the supercycle orbit of $f^{p*2^{n}}$, that is, $CI_{1}.....I_{p*2^{n}-1}$.

ii) The function $f^{p*2^{n}}$can be approximated by a quadratic
polynomial $g(x)$ in the neighborhood of each point of the supercycle
$\left\{ x_{1},...x_{p*2^{n}}\right\} $ \cite{Hunt99}.

The period-$p*(q*2^{n})$ saddle-node orbit of $f$ occurs when $g(x)$
has a period-$q$ saddle-node orbit. This period-$q$ saddle-node
orbit is like $\left\{ x_{0},g(x_{0}),.....,g^{q-1}(x_{0})\right\} $,
with $g^{q}(x_{0})=x_{0}$. As $g(x)$ is an approximation of $f^{p*2^{n}}$to
pass from point $g^{i}(x_{0})$ to point $g^{i+1}(x_{0})$ we will
have to iterate $p*2^{n}$ times the function $f$, that is, the iterated
of $f$ visit the neighborhood of the points of the supercycle $\left\{ x_{1},...x_{p*2^{n}}\right\} $.
With each $p*2^{n}$iterated, the same neighborhood is visited again,
not to the same point but the following one of the period-$q$ saddle-node
orbit trapped in this neighborhood.

With each $p*2^{n}$ iterated of $f$ we go back to the same neighborhood
of a given point of the supercycle. This neighborhood will be to the
left or to the right of the center $C$, or it will be the point $C$
itself. For the time being, let's leave the neighborhood of the point
$C$ and let's consider the period-$q$ saddle-node orbits trapped
in the other neighborhoods. If this orbit is trapped in a neighborhood
to the left (right) of the center $C$ then all its points will be
placed to the left (right) of $C$ and they will generate an L (R)
in the sequence of the period-$p*(q*2^{n})$ saddle-node orbit. Therefore,
writing consecutively $q$ times the sequence of the orbit of the
supercycle we would obtain the sequence of the period-$p*(q*2^{n})$
orbit.

iii) A drawback of the previous procedure takes place when the neighborhood
of the point $C$ is visited. For in this neighborhood not all the
points of the period-$q$ saddle-node orbit are to the left or to
the right of the point $C$. Apparently, to solve the problem, it
would be enough to calculate the sequence of the period-$q$ saddle-node
orbit, and with this sequence to replace the $C_{1}C_{2}.....C_{q}$
of the sequence $C_{1}I_{1}.....I_{p*2^{n}-1}C_{2}I_{1}.....I_{p*2^{n}-1}.........C_{q}I_{1}.....I_{p*2^{n}-1}$.
This is so because each time the neighborhood of $C$ is visited a
point of the period-$q$ saddle-node orbit trapped in this neighborhood
is visited.

iv) However there is a drawback as the letters $C_{1}C_{2}.....C_{q}$are
replaced by the the letters of period-$q$ saddle-node orbit, because
the period-$q$ saddle-node orbit is conjugated in the neighborhood
of $C$, as the value of $n$ is in $f^{p*2^{n}}$. This is so because
of the negative sign in Eq. (\ref{15 nueva 20}). If $n$ is even
the sequence holds and if $n$ is odd the sequence is conjugated.
Therefore to solve the problem $\left(-1\right)^{n}$ is calculated
and if this value is negative the sequence of the orbit is conjugated.

v) Finally, it is enough to replace the letters $C_{1}C_{2}.....C_{q}$
by the letters of the period-$q$ saddle-node orbit sequence correctly
conjugated to obtain the right result.

Notice that it is enough to know the sequences of the supercycles
of $f^{2^{n}}$and the primary saddle-node orbits $f^{p}$ to obtain
the sequence of each orbit in the logistic map. According to the rule
that we show here we get the sequence of the supercycle of $f^{p*2^{n}}$in
the period-$p$ window, and from this supercycle we calculate the
sequence of the period-$p*(q*2^{n})$ saddle-node orbit that will
occur in a period-$q$ window inside a period-$p$ window.

Let's see how the principle is used to obtain the period-$3*2$ saddle-node
orbit sequence of Fig. \ref{cap:fig2}, in which $p=1,n=1,q=3$.

i) This time there is a supercycle of $f^{2}$, whose orbit is formed
by the points $\left\{ \frac{1}{2},f(\frac{1}{2})\right\} $and its
own sequence is $CR$.

ii) There is a period-$3$ saddle-node orbit which is trapped in the
neighborhoods of $\frac{1}{2}$ and $f(\frac{1}{2})$, accordingly
we write $CRCRCR$

iii) The sequence of the period-$3$ saddle-node is $RRL$, as shown
in Fig. \ref{cap:fig1}

iv) As $\left(-1\right)^{1}=-1$ the sequence $RRL$ is conjugated
to give $LLR$

v) Finally, the letters $C$ of the sequence $CRCRCR$ are replaced
by the letters of the orbit $LLR$, turning into $LRLRRR$. The first
letter indicates a place with respect to $C$, the point $x=\frac{1}{2}$.
In Fig. \ref{cap:fig2} it can be observed that the saddle-node orbit
goes through as the sequence obtained .

The principle of ordering shown above, for saddle-node orbits, is
very similar to the one described by Derrida \emph{et al.} in \cite{Derrida79}
for supercycle orbits.

\textbf{Principle of saddle-node orbits} \textbf{ordering}.- Let be
$f$ the logistic map, if $q>s$ then the period-$qp2^{n}$saddle-node
orbit occurs at a value $r$ smaller than that for which the period-$sp2^{n}$saddle-node
orbit occurs, that is, the period $qp2^{n}$ precedes period $sp2^{n}$
in the parameter space. We note it down like $qp2^{n}\triangleright sp2^{n}$
if $q>s$.

Proof

As the function $f^{p*2^{n}}$can be approximated by a quadratic polynomial
$g(x)$ in the neighborhood of each point of the supercycle $\left\{ x_{1},...x_{p*2^{n}}\right\} $
\cite{Hunt99}, it will turn out that as $g(x)$ has a period-$s$
saddle-node orbit then $f$ will have a period-$sp2^{n}$ orbit ,
and as $g(x)$ has a period-$q$ saddle-node orbit then $f$ will
have a period-$qp2^{n}$ orbit. Since for the saddle-node $q\triangleright s$
if $q>s$ so $qp2^{n}\triangleright sp2^{n}$ if $q>s$.

Comments:

i) Although we have a Sharkovsky ordering the principle is not limited
to first-occurrence orbits to which the Sharkovsky theorem applies.

ii) We cannot discard halfway orbits in this ordering, which come
from other values of $n$, $p$ or $q$, that is, we have ordered
non-first-occurrence orbits althugh all of them.

\section{Conclusion}

The Feigenbaum equation

\[
r_{n+1}=\frac{1}{\delta}r_{n}+(1-\frac{1}{\delta})r_{\infty}\]
which allows us to locate the successive pitchfork bifurcations in
a period doubling cascade, turns out to be an expression of a general
nature. This is applied to cascades of other objects present in the
logistic equation. For instance, this expression rules, among other
things, saddle-node bifurcations, Myrberg-Feigenbaum points, and the
birth and death of some windows. That is, not only the equation gives
the ordering but also the parameter values for which such objects
appear.

An immediate consequence of these results is the posibility to determine:
the scaling of windows, the parameters of birth and death, the relative
position of the Myrberg-Feigenbaum points, the relative position of
the attractors in which the period doubling cascade finishes.

We would like to highlight another consequence of these results, a
new concept: an attractor of attractors, which coincide with the Myrberg-Feigenbaum
points.

The former results have been first expounded with linear transformations
benefiting from the properties of the windows (bifurcation rigidity
Principle), and later, part of theses results have been derived using
the linearization of the Feigenbaum-Cvitanoviç equation. Later the
bifurcation rigidity Principle is generalized to functions with a
maximum other than quadratic.

The concept upon which the whole work relies is the saddle-node bifurcation
cascade, which is a sequence of successive saddle-node bifurcations,
which in turn double the number of saddle-node fixed points. The bifurcation
points are ruled by the same equation obtained by Feigenbaum for the
period doubling cascade. By using the bifurcation parameter given
by saddle-node bifurcation cascade, we follow the Feigenbaum work
to end up in a Feigenbaum-Cvitanoviç equation, which proves the universal
behavior of our results, and it allows us the generalization to maps
other than the logistic one. 

The saddle-node bifurcation cascade shows a further behavior similar
to a period doubling cascade: it is self-similar after successive
iterations. Nonetheless, for a fixed period that is duplicated in
each bifurcation, there are different saddle-node bifurcations, however
this they have the same accumulating point and the same scaling law.

The ordering of orbits missing from Sharkovsky theorem have been given,
as well as the ordering of Myrberg-Feigenbaum points associated to
the windows born of saddle-node bifurcation cascades.

Lastly symbolic sequences of the orbits found in the saddle-node bifurcation
cascade have been described. 

We have summarized above the mathematical contributions of our work.
We started the introduction by pointing out that any progress, in
the logistic equation, would also yield progress to Physics . We would
like to quote {}``\textbf{the self-similar cascade of bifurcations}
(....) \textbf{it is characterized by an infinite series of saddle-node
bifurcation of cycles, accumulating at a finite parameter value}''
reported by Yeung and Strogatz \cite{Yeung00}. This mechanism coincides
with the one described for a saddle-node bifurcation cascade, where
the accumulating parameter value is the Myrberg-Feigenbaum point,
which will most likely make the phenomenon fit in the framework described
in this work.

\section*{Acknowledgments}

The author wishes to thank Daniel Rodríguez-Pérez for helpful discussions
and help in the preparation of the manuscript.

\newpage

\begin{table}[h]

\caption{\label{cap:table1}Period $3\cdot2^{n}$ Saddle-Node bifurcation
cascade inside the canonical window. The convergence to $\alpha$
and $\delta$ Feigenbaum constants is shown.}

\begin{center}\begin{tabular}{|c|c|c|c|c|}
\hline 
Period&
$r_{3*2^{n}}$&
$\delta_{3*2^{n}}$&
$\alpha_{3*2^{n}}$&
$n$\tabularnewline
\hline
\hline 
$3$&
3.82842712&
&
&
0\tabularnewline
\hline 
$3*2$&
3.62655316&
&
2.30066966&
1\tabularnewline
\hline 
$3*2^{2}$&
3.58202300&
4.5334294&
2.52112183&
2\tabularnewline
\hline 
$3*2^{3}$&
3.57253281&
4.69223061&
2.49753642&
3\tabularnewline
\hline 
$3*2^{4}$&
3.57049971&
4.66784221&
2.50264481&
4\tabularnewline
\hline 
$3*2^{5}$&
3.57006433&
4.66971381&
2.50263906&
5\tabularnewline
\hline
\end{tabular}\end{center}
\end{table}

\begin{table}[h]

\caption{\label{cap:table2}Period $5\cdot2^{n}$ Saddle-Node bifurcation
cascade inside the canonical window. The convergence to Feigenbaum
$\delta$ constant is shown.}

\begin{center}\begin{tabular}{|c|c|c|c|}
\hline 
Period&
$r_{5*2^{n}}$&
$\delta_{5*2^{n}}$&
n\tabularnewline
\hline
\hline 
$5$&
3.73817238&
&
0\tabularnewline
\hline 
$5*2$&
3.60520807&
&
1\tabularnewline
\hline 
$5*2^{2}$&
3.57751225&
4.80088006&
2\tabularnewline
\hline 
$5*2^{3}$&
3.57156559&
4.65737406&
3\tabularnewline
\hline 
$5*2^{4}$&
3.57029262&
4.67144810&
4\tabularnewline
\hline 
$5*2^{5}$&
3.57001998&
4.66925870&
5\tabularnewline
\hline
\end{tabular}\end{center}
\end{table}

\begin{table}[h]

\caption{\label{cap:table3}Period $3\cdot2^{n}$ Saddle-Node bifurcation
cascade inside the period-3 window. The convergence to Feigenbaum
$\delta$ constants is shown.}

\begin{center}\begin{tabular}{|c|c|c|c|}
\hline 
Period&
$r_{3*2^{n}}$&
$\delta_{53*2^{n}}$&
n\tabularnewline
\hline
\hline 
$3$&
3.85361311&
&
0\tabularnewline
\hline 
$3*2$&
3.85031470&
&
1\tabularnewline
\hline 
$3*2^{2}$&
3.84962047&
4.75117756&
2\tabularnewline
\hline 
$3*2^{3}$&
3.84947362&
4.72747702&
3\tabularnewline
\hline 
$3*2^{4}$&
3.84944223&
4.67824148&
4\tabularnewline
\hline 
$3*2^{5}$&
3.84943551&
4.67113095&
5\tabularnewline
\hline
\end{tabular}\end{center}
\end{table}

\begin{table}[h]

\caption{\label{cap:table4}Period $3\cdot2^{n}$ Saddle-Node bifurcation
cascade inside the period-5 window. The convergence to Feigenbaum
$\delta$ constants is shown.}

\begin{center}\begin{tabular}{|c|c|c|c|}
\hline 
Period&
$r_{3*2^{n}}$&
$\delta_{3*2^{n}}$&
n\tabularnewline
\hline
\hline 
$3$&
3.744003&
&
0\tabularnewline
\hline 
$3*2$&
3.74321655&
&
1\tabularnewline
\hline 
$3*2^{2}$&
3.743050198&
4.727522&
2\tabularnewline
\hline 
$3*2^{3}$&
3.7430150831&
4.737633&
3\tabularnewline
\hline 
$3*2^{4}$&
3.7430075758&
4.677634&
4\tabularnewline
\hline 
$3*2^{5}$&
3.7430059688&
4.671437&
5\tabularnewline
\hline
\end{tabular}\end{center}
\end{table}

\newpage

\newpage

\newpage

\newpage

\newpage

\begin{figure}[h]

\caption{\label{cap:fig5}Period-$3\cdot2$ Saddle-Node orbit in the period-$5$
window, for $r\simeq3.74321655$. The shape of Fig. \ref{cap:fig2}
is reproduced five times; here is shown a magnification of the replica
near $(1/2,\,1/2)$.}

\includegraphics[%
  width=0.70\textwidth,
  keepaspectratio,
  angle=-90]{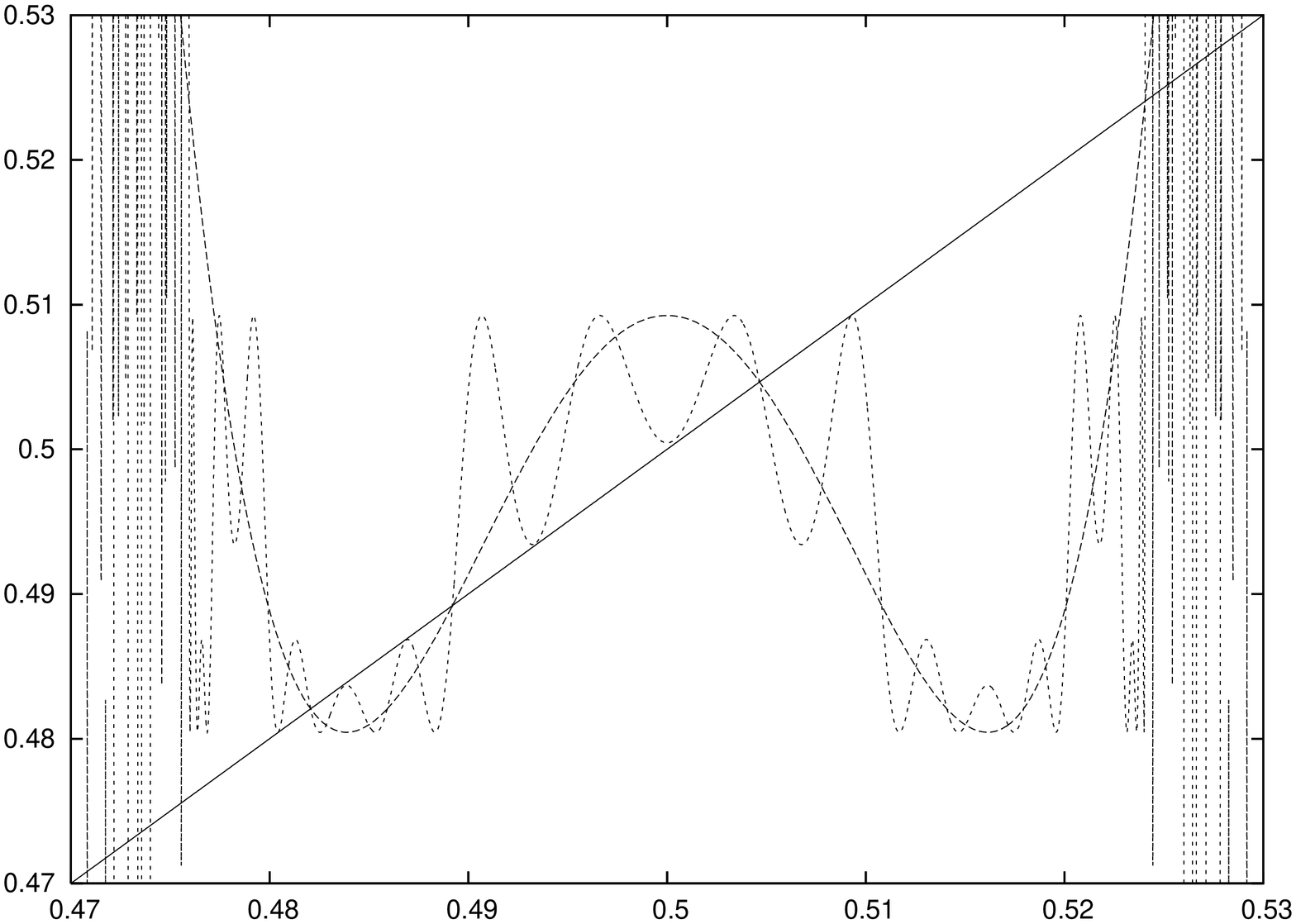}
\end{figure}

\newpage

\begin{figure}[h]

\caption{\label{cap:fig6}One of the three possible period-$6$ Saddle-Node
orbits close to $r=3.97790000$.}

\includegraphics[%
  width=0.70\textwidth,
  keepaspectratio,
  angle=-90]{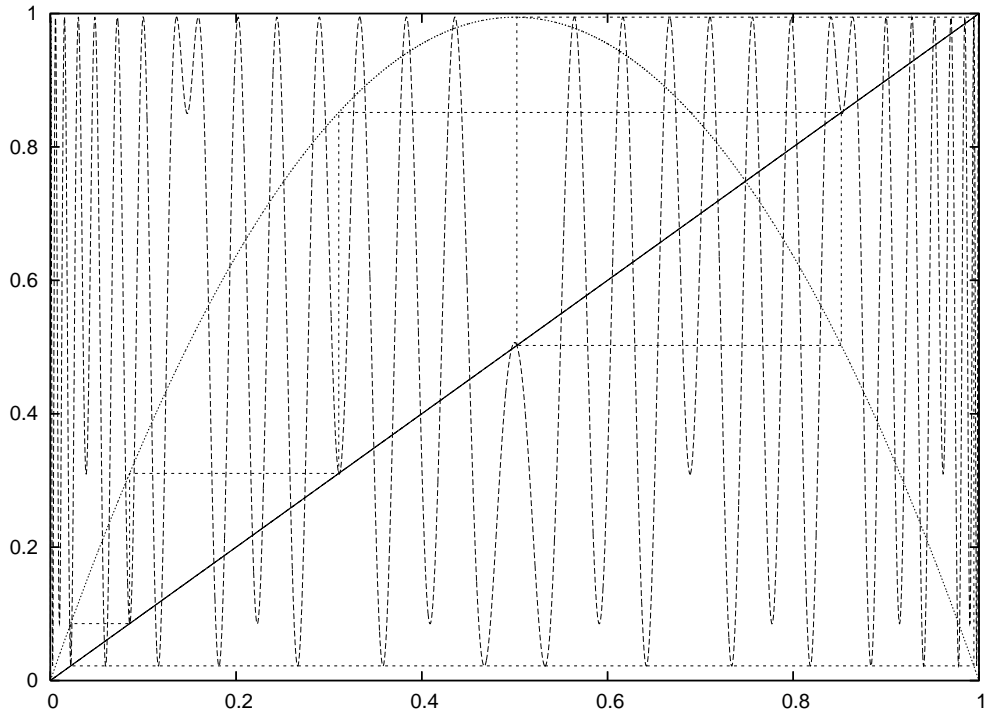}
\end{figure}

\newpage
\begin{figure}[h]

\caption{\label{cap:fig7}Period-$6\cdot2$ Saddle-Node orbit in the canonical
window, close to $r\simeq3.66840000$. The shape of Fig. \ref{cap:fig6}
is reproduced twice.}

\includegraphics[%
  width=0.70\textwidth,
  keepaspectratio,
  angle=-90]{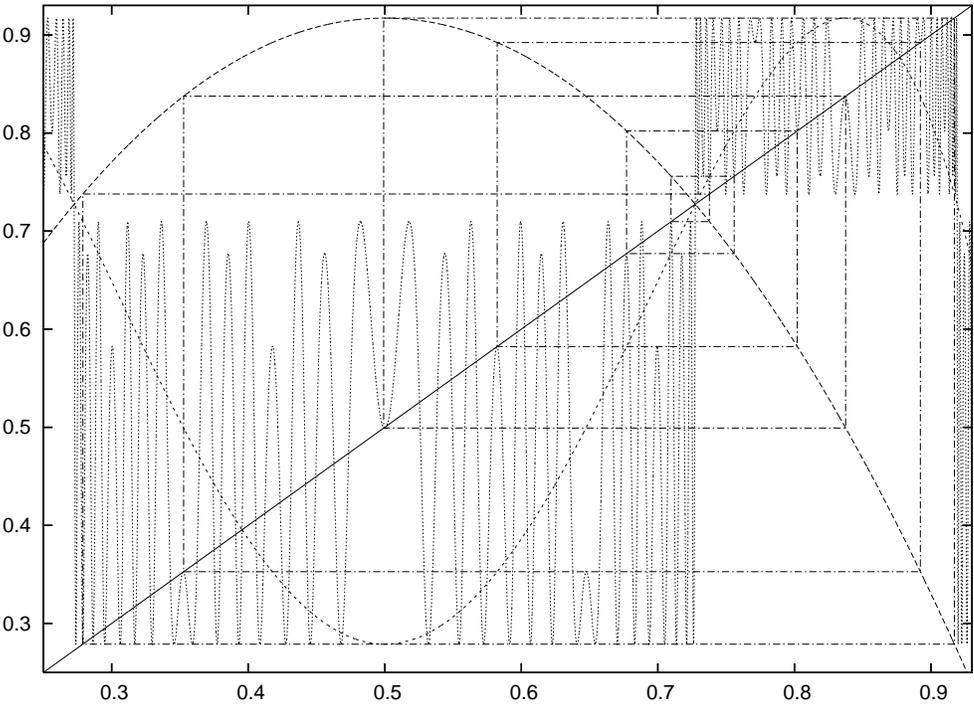}
\end{figure}

\end{document}